\documentclass[twocolumn,showpacs,preprintnumbers,amsmath,amssymb]{revtex4}
\usepackage{graphicx}
\usepackage{dcolumn}
\usepackage{bm}

\newcommand{\bi}{\begin{itemize}}
\newcommand{\ei}{\end{itemize}}
\newcommand{\be}{\begin{equation}}
\newcommand{\ee}{\end{equation}}
\newcommand{\ba}{\begin{eqnarray}}
\newcommand{\ea}{\end{eqnarray}}
\newcommand{\bse}{\begin{subequations}}
\newcommand{\ese}{\end{subequations}}

\newcommand{\CS}{{\cal{S}}}
\newcommand{\la}{\langle}
\newcommand{\ra}{\rangle}
\newcommand{\kB}{k_{_B}}
\newcommand{\vrot}{v_{\hbox{\tiny{rot}}}}

\newcommand{\rhohac}{\rho_c^{\hbox{\tiny{(h)}}}}
\newcommand{\nha}{n^{\hbox{\tiny{(h)}}}}

\newcommand{\sha}{s^{\hbox{\tiny{(h)}}}}
\newcommand{\shac}{s_c^{\hbox{\tiny{(h)}}}}
\newcommand{\shacth}{s_c^{\hbox{\tiny{(h)}}}|_{_{\hbox{\tiny{th}}}}}
\newcommand{\shacmb}{s_c^{\hbox{\tiny{(h)}}}|_{_{\hbox{\tiny{MB}}}}}
\newcommand{\shacem}{s_c^{\hbox{\tiny{(h)}}}|_{_{\hbox{\tiny{em}}}}}
\newcommand{\nhac}{n_c^{\hbox{\tiny{(h)}}}}
\newcommand{\xhac}{x_c^{\hbox{\tiny{(h)}}}}
\newcommand{\Tha}{T^{\hbox{\tiny{(h)}}}}

\newcommand{\xha}{x^{\hbox{\tiny{(h)}}}}
\newcommand{\sigha}{\sigma_{\hbox{\tiny{(h)}}}}

\newcommand{\neqq}{n^{\hbox{\tiny{(eq)}}}}

\newcommand{\RT}{R_{\hbox{\tiny{tot}}}}
\newcommand{\RC}{R_{\hbox{\tiny{core}}}}
\newcommand{\xf}{x_{\hbox{f}}}
\newcommand{\nf}{n_{\hbox{f}}}
\newcommand{\sff}{s_{\hbox{f}}}

\begin{document}

\title{Using Entropy to Discriminate Annihilation Channels in 
Neutralinos Making up Galactic Halos} 

\author{Luis G Cabral--Rosetti$^\dagger$, Xavier Hern\'andez$^\ddagger$ and
Roberto A Sussman$^\dagger$}


\affiliation{$^\dagger$Instituto de Ciencias Nucleares,  
Universidad Nacional Aut\'onoma de M\'exico \\
A. P. 70--543,  M\'exico 04510 D.F., M\'exico \\ 
$^\ddagger$Instituto de Astronom\'\i a,  
Universidad Nacional Aut\'onoma de M\'exico \\
A. P. 70--543,  M\'exico 04510 D.F., M\'exico}


\begin{abstract}
Applying the microcanonical definition of entropy to a weakly interacting and
self--gravitating neutralino gas, we evaluate the change in the local entropy 
per particle of this gas between the freeze out era and present day virialized 
halo structures. An ``entropy consistency'' criterion emerges by comparing 
theoretical and empirical estimates of this entropy.  We apply this criterion 
to the cases when neutralinos are mostly B-inos and mostly Higgsinos, in 
conjunction with the usual ``abundance'' criterion requiring that present 
neutralino  relic density complies with $0.2<\Omega_{{\tilde\chi^1_0}} < 0.4$ 
for $h\simeq 0.65$.  The joint application of both criteria reveals that a 
much better fitting occurs for the B-ino than for the Higgsino channels, 
so that the former seems to be a favored channel along the mass range  of
$155\,\hbox{GeV} <  m_{{\tilde\chi^1_0}} < 230 \,\hbox{GeV}$. These results 
are consistent with neutralino annihilation patterns that emerge from recent 
theoretical analysis on cosmic ray positron excess data reported by the HEAT 
collaboration. The suggested methodology can be applied to test other 
annihilation channels of the neutralino, as well as other particle candidates 
of thermal WIMP gas relics.                     
\end{abstract}


\maketitle

\section{Introduction}

There are strong theoretical arguments favoring lightest supersymmetric 
particles (LSP) as making up the relic gas that forms the halos of actual 
galactic structures. Assuming that  {\it R} parity is conserved and that 
the LSP is stable, it might be an ideal candidate for cold dark matter 
(CDM), provided it is neutral and has no strong interactions. The most 
favored scenario \cite{Ellis,Report,Torrente,Roszkowski,Fornengo,Ellis2} 
considers the LSP to be the lightest neutralino ($\tilde\chi_1^0$), a 
mixture of supersymmetric partners of the photon, $Z$ boson and neutral 
Higgs boson \cite{Report}. Since neutralinos must have decoupled once 
they were non-relativistic, it is reasonable to assume that they constituted 
originally a Maxwell-Boltzmann (MB) gas in thermal equilibrium with other
components of the primordial cosmic plasma. In the present cosmic
era, such a gas is practically collision--less and is either virialized 
in galactic and galactic cluster halos, in the process of virialization 
or still in the linear regime for superclusters and structures near the 
scale of homogeneity\cite{KoTu, Padma1,Peac}.

Besides the constraint due to their present abundance as main constituents 
of cosmic dark matter ($\Omega_{{\tilde\chi^1_0}} \sim 0.3$), it is still 
uncertain which type of annihilation cross section characterizes these 
neutralinos. In this paper we present a method that discriminates between 
different cross sections, based on demanding that (besides yielding the 
correct abundance) a theoretically  estimated entropy per particle matches 
an empiric  estimate of the same entropy, both constructed for actual 
galactic dark halo structures. The application of this ``entropy consistency''
criterion is straightforward because entropy is a state variable that can 
be evaluated at equilibrium states, irrespective of how enormously 
complicated the evolution between each such state might have been. In this 
context, the two fiducial equilibrium states of the neutralino gas are 
(to a good approximation) the decoupling (or ``freeze out'') epoque and 
their present state as a virialized relic gas. Considering simplified forms 
of annihilation cross sections, the joint application of the abundance 
and entropy--consistency criteria favors the neutralinos as mainly ``B--inos''
over neutralinos as mainly ``higgsinos''. These results are consistent with 
the theoretical analysis of the HEAT experiment~\cite{HEAT-TH,HEAT1,HEAT2} 
which aims at relating the observed positron excess in cosmic rays with 
a possible weak interaction between neutralinos and nucleons in galactic 
halos.
 
The paper is organized as follows. In section 2 we describe the thermodynamics
of the neutralino gas as it decouples. The microcanonical ensemble entropy is 
applied in section 3 to the post--decoupling neutralino gas to estimate the 
change in entropy between freeze out and present day conditions, leading to 
a suitable theoretical estimate of the entropy per particle. In section 4 we 
obtain an empirical estimate of this entropy based on actual halo variables, 
while in section 5 we examine the consequences of demanding that these two 
entropies coincide. We summarize our results in section 6.    

\section{The Neutralino Gas}

Before the freeze out, the neutralinos satisfy the thermal equation of state 
of a non-relativistic MB gas \cite{KoTu,Padma1,Peac}
%
\ba \rho \ =&& \
m_{{\tilde\chi_0^1}}\,n_{{\tilde\chi_0^1}}\,
\left(1+\frac{3}{2\,x}\right),\qquad p \ = \
\frac{m_{{\tilde\chi_0^1}}\,n_{{\tilde\chi_0^1}}}{x}, \label{eqst}\\
 x \ \equiv&& \
\frac{m_{{\tilde\chi_0^1}}}{T},\label{beta_def}\ea
%
where $m_{{\tilde\chi_0^1}}$ and $n_{{\tilde\chi_0^1}}$ are the neutralino
mass and number density. Since we will deal exclusively with the lightest 
neutralino, we will omit henceforth the subscript $_{{\tilde\chi_0^1}}$, 
understanding that all usage of the term ``neutralino'' and all symbols of 
physical and observational variables ({\it{i.e.}} $\Omega_0,\,m,\,\rho,\,n,$
etc.) will correspond to this specific particle. As long as the neutralino 
gas is in thermal equilibrium, we have
\ba n \ \approx \ \neqq \ =&& \
g\,\left[\frac{m}{\sqrt{2\,\pi}}\right]^3\,x^{-3/2}\,\exp\,
\left(-x\right),\label{n_theq}
\ea        
where $g=2$ is the degeneracy factor of the neutralino species. The number 
density $n$ satisfies the Boltzmann equation \cite{Report,KoTu}
\ba \dot n + 3\,H\,n \ = \ -\la
\sigma|\hbox{v}|\ra\left[n^2-\left(\neqq\right)^2\right],\label{boltz}\ea
where $H$ is the Hubble expansion factor and $\la \sigma|\hbox{v}|\ra$ is 
the annihilation cross section.  Since the neutralino is non-relativistic as
annihilation reactions ``freeze out'' and it decouples from the radiation 
dominated cosmic plasma, we can assume for $H$ and $\la\sigma|\hbox{v}|\ra$ 
the following forms
\ba H \ = \ 1.66\, g_*^{1/2}\frac{T^2}{m_p},\label{eqH}\\
\la\sigma|\hbox{v}|\ra \ = \ a \ + \ b\la \hbox{v}^2\ra,\label{eq<sv>}\ea 
where $m_p=1.22\times 10^{19}$ GeV is Planck's mass, $g_*=g_*(T)$ is the 
sum of relativistic degrees of freedom, $\la \hbox{v}^2\ra$ is the thermal 
averaging of the center of mass velocity (roughly $\hbox{v}^2\propto 1/x$ 
in non-relativistic conditions) and the constants $a$ and $b$ are determined 
by the parameters characterizing specific annihilation processes of the
neutralino (s-wave or p-wave) \cite{Report}. The decoupling of the
neutralino gas follows from the condition
\ba \Gamma \ \equiv \ n\,\la\sigma|\hbox{v}|\ra \ = \ H,\label{fcond}\ea
leading to the freeze out temperature $T_{\hbox{f}}$. Reasonable approximated 
solutions of (\ref{fcond}) follow by solving for $x_f$ the implicit relation 
\cite{Report}
\ba \xf  = 
\ln\left[\frac{0.0764\,m_p\,c_0(2+c_0)\,(a+6\,b/\xf)\,m}{(g_{*{\hbox{f}}}
\,\xf)^{1/2}}\right],
\label{eqxf}\ea  
where $g_{*{\hbox{f}}}=g_*(T_{\hbox{f}}) $ and $c_0\approx 1/2$ yields the 
best fit to the numerical solution of (\ref{boltz}) and (\ref{fcond}). From 
the asymptotic solution of (\ref{boltz}) we obtain the present abundance of 
the relic neutralino gas \cite{Report}
%
\ba \Omega_0\,h^2 \ = &&\ Y_\infty\, \frac{\CS_0\, m} {\rho_{\hbox{crit}}/h^2}
\ \approx \ 2.82\times 10^8\,Y_\infty\,\frac{m}{\hbox{GeV}},
\label{eqOmega0}
\ea
where
\ba
 Y_\infty \
\equiv  &&\ \frac{n_0}{\CS_0} \nonumber\\ = && 
\left[0.264\,g_{*\hbox{f}}^{1/2}\,m_p\,m\left\{a/\xf+3(b-1/4\,a)
/\xf^2\right\}\right]^{-1},
\nonumber
\\
\label{eqYinf}\ea
%
and $\CS_0\approx 4000\,\hbox{cm}^{-3}$ is the present radiation entropy
density (CMB plus neutrinos),  $\rho_{\hbox{crit}} = 1.05 \times
10^{-5}\,\hbox{GeV}\,\hbox{cm}^{-3}$. 

Since neutralino masses are expected to be in the range of tens to hundreds 
of GeV's and typically we have $\xf\sim 20$ so that $T_{\hbox{f}} >$ GeV, we
can use $g_{*{\hbox{f}}}\simeq 80$ \cite{Report} in equations (\ref{eqxf}) --
(\ref{eqYinf}). Equation (\ref{eqxf}) shows how $\xf$ has a logarithmic
dependence on $m$, while theoretical considerations 
\cite{Ellis,Report,Torrente,Roszkowski,Fornengo,Ellis2} related to the minimal
supersymetric extensions of the Standard Model (MSSM) yield specific forms 
for $a$ and $b$ that also depend on $m$. Inserting into 
(\ref{eqOmega0})--(\ref{eqYinf}) the specific forms of $a$ and $b$ for each 
annihilation channel leads to a specific range of $m$ that satisfies the 
``abundance'' criterion based on current observational constraints that 
require $0.2 < \Omega_0 < 0.4$ and $h\approx 0.65$ \cite{Peac}.

Suitable forms for $\la \sigma |v|\ra$ can be obtained for all types of 
annihilation reactions \cite{Report}. If the neutralino is mainly pure B-ino, 
it will mostly annihilate into lepton pairs through t-channel exchange of 
right-handed sleptons. In this case the cross section is p-wave dominated 
and can be approximated by (\ref{eq<sv>}) with \cite{Torrente,Moroi,Olive}
\ba a \ \approx \ 0,\qquad b \ \approx
\ \frac{8\,\pi\,\alpha_1^2}{m^2\,\left[1+m_l^2/m^2\right]^2},
\label{sleptons}
\ea
where $m_{l}$ is the mass of the right-handed slepton ($m_{l} \sim m$ 
\cite{Torrente}) and $\alpha_1 = g_1^2/4 \pi \simeq 0.01$ is the fine 
structure coupling constant for the $U(1)_Y$ gauge interaction. If the 
neutralino is Higgsino-like, annihilating  into W-boson pairs, then the 
cross section is s-wave dominated and can be approximated by (\ref{eq<sv>}) 
with \cite{Torrente,Moroi,Olive}
\ba b \ \approx \ 0,\qquad a \ \approx \
\frac{\pi\,\alpha_2^2\,(1-m_{ W}^2/m^2)^{3/2}}{2\,m^2\,(2-m_{_W}^2/m^2)^2},
\label{Wboson}
\ea
where $m_{_W}=80.44$ GeV is the mass of the W-boson and 
$\alpha_2 = g_2^2/4 \pi \simeq 0.03$ is the fine structure coupling constant 
for the $SU(2)_L$ gauge interaction.      

In the freeze out era the entropy per particle (in units of the Boltzmann 
constant $\kB$) for the neutralino gas is given by \cite{KoTu,Peac,Padma1}
\ba \sff \ = \ \left[\frac{\rho + p}{n\,T}\right]_{\hbox{f}} \ = \
\frac{5}{2} \ + \ \xf,\label{sf}\ea
where we have assumed that chemical potential is negligible and have used
the equation of state (\ref{eqst}). From (\ref{eqxf}) and (\ref{sf}), it is 
evident that the dependence of $\sff$ on $m$ will be determined by the 
specific details of the annihilation processes through the forms of $a$ and 
$b$. In particular, we will use (\ref{sleptons}) and (\ref{Wboson}) to 
compute $\sff$ from (\ref{eqxf})-(\ref{sf}).

\section{The Microcanonical Entropy}

After the freeze out era, particle numbers are conserved and the neutralinos 
constitute a weakly interacting and practically collision--less and 
self--gravitating gas. This gas is only gravitationally coupled to all other 
components of the cosmic fluid. As it expands,  it experiences free streaming 
and eventually undergoes gravitational clustering forming stable bound 
virialized structures \cite{Peac,Padma1,Padma2,Padma3}. The evolution between 
a spectrum of density perturbations at the freeze out and the final virialized
structures is extremely complex, involving a variety of dissipative effects 
characterized by collisional and collision--less relaxation processes 
\cite{Padma2,Padma3,HD}.  However, the freeze out and present day virialized 
structures roughly correspond to ``initial'' and ``final'' equilibrium states 
of this gas. Therefore, instead of dealing with the enormous complexity of 
the details of the intermediary processes, we will deal only with quantities 
defined in these states with the help of simplifying but general physical 
assumptions. 

While the assumption of an ideal (thermodynamical) gas complying with 
Maxwell-Boltzmann statistics is justified before the freeze out (because the 
collision--dominated interactions are short--ranged), this type of 
thermodynamical formalism, associated with extensive entropy and energy and 
with a well defined thermodynamical limit, cannot be applied to present day 
collision--less neutralinos subject to a long--range gravitational interaction
\cite{Padma2,Padma3}. In these conditions, the appropriate approach follows 
from the microcanonical ensemble in the ``mean field'' approximation that 
yields an entropy definition that is well defined for a self--gravitating gas 
in an intermediate scale, between the short range and long range regimes of 
the gravitational potential. This intermediate scale can be associated with 
a region that is ``sufficiently large as to contain a large number of 
particles but small enough for the gravitational potential to be treated as 
a constant''~\cite{Padma2}. Considering the neutralino gas in present day 
virialized halo structures as a diluted, non-relativistic (nearly) ideal 
gas of weakly interacting particles, its microcanonical entropy per particle 
under these conditions can be given in terms of the volume of phase space 
\cite{Padma3} 
\ba 
s \ = \ \ln \,\left[\frac{\,(2mE)^{3/2}\,V\,}{(2\pi\hbar)^3}\right],
\label{mcsdef}
\ea
where $V$ and $E$ are local average values of volume and energy associated 
with the intermediate scale. For  non-relativistic velocities $v/c\ll 1$, 
we have $V\propto 1/n\propto m/\rho$ and $E\propto m\,v^2/2\propto m/x$. 
In fact, since the microcanonical description is more fundamental, definition
(\ref{mcsdef}), evaluated at the freeze out, is consistent with (\ref{n_theq}) 
and (\ref{sf}), and so it is also valid immediately after the freeze out era 
(once particle numbers are conserved). Since (\ref{mcsdef}) is valid at both 
the initial and final states, respectively corresponding to the freeze out 
($\sff,\,\xf,\,\nf$) and to the values associated with a suitable halo 
structure ($\sha,\,\xha,\,\nha$), the change in entropy per particle that 
follows from (\ref{mcsdef}) between these two states is given by
\ba \sha \ - \ \sff \ = \ 
\ln\,\left[\frac{\nf}{\nha}\left(\frac{\xf}{\xha}\right)^{3/2}
\right],\label{Delta_s}\ea
where (\ref{sf}) can be used to eliminate $\sff$ in terms of $\xf$. 
Considering present day halo structures as roughly spherical, inhomogeneous 
and self-gravitating gaseous systems, it is safe to assume that near the 
symmetry center the spacial gradients of all macroscopic quantities are 
negligible~\cite{HG,IS2}. Even if one assumes a ``cuspy'' density profile, 
as predicted by numerical simulations, the gravitational potential of the 
halo can be considered to be practically constant within any typical region 
of up to $\sim 10\,\hbox{pc}^3$ within the halo core (a scale that is smaller 
than the resolution scale of these simulations~\cite{NSres}). Since such 
a typical region fits reasonably well the conditions associated with the 
intermediate scale of the microcanonical description, we will consider 
current halo macroscopic variables as evaluated near the center of the halo:
$\shac,\,\xhac,\,\nhac$.      

In order to obtain a convenient theoretical estimate of $\shac$ from 
(\ref{Delta_s}), we need to relate $\nf$ with present day cosmological 
parameters like $\Omega_0$ and $h$. Bearing in mind that density 
perturbations at the freeze out era were very small 
($\delta\,\nf/\nf < 10^{-4}$, \cite{KoTu,Padma1,Peac}), the density $\nf$ 
is  practically homogeneous and so we can estimate it from the conservation 
of particle numbers: $\nf = n_0\,(1+z_{\hbox{f}})^3$, and of photon entropy: 
$g_{*\hbox{f}}\CS_{\hbox{f}} = g_{*0}\,\CS_0\, \,(1+z_{\hbox{f}})^3$, 
valid from the freeze out era to the present for the unperturbed homogeneous 
background. Eliminating $ (1+z_{\hbox{f}})^3$ from these conservation laws 
yields
%
\ba \nf \ = \
n_0\,\frac{g_{*\hbox{f}}}{g_{*0}}\left[\frac{T_{\hbox{f}}}
{T_0^{\hbox{\tiny{CMB}}}}\right]^3
\ \simeq \ 20.46\,n_0\,
\left[\frac{x_0^{\hbox{\tiny{CMB}}}}{\xf}\right]^3,\label{eqnf}\\
\hbox{where}\quad x_0^{\hbox{\tiny{CMB}}} \ \equiv \
\frac{m}{T_0^{\hbox{\tiny{CMB}}}}
\ = \ 4.29\,\times\,10^{12}\,\frac{m}{\hbox{GeV}}\nonumber\ea
%
and $g_{*0}=g_*(T_0^{\hbox{\tiny{CMB}}})\simeq 3.91$ and
$T_0^{\hbox{\tiny{CMB}}}=2.7\,\hbox{K}$. Since for present day conditions 
$n_0/\nhac=\rho_0/\rhohac$ and $\rho_0=\rho_{\hbox{crit}}\,\Omega_0\,h^2 $, 
we collect the results from (\ref{eqnf}) and write (\ref{Delta_s}) as 
\ba \shacth = \xf + 92.78 +
\ln\left[\left(\frac{m}{\hbox{GeV}}\right)^3\,\frac{h^2\,\Omega_0}
{(\xf\,\xhac)^{3/2}}\,\frac{\rho_{\hbox{crit}}}{\rhohac}\right]\nonumber\\
= \xf + 81.31 +
\ln\left[\left(\frac{m}{\hbox{GeV}}\right)^3\,\frac{h^2\,\Omega_0}
{(\xf\,\xhac)^{3/2}}\,\frac{\hbox{GeV/cm}^3}{\rhohac}\right]\nonumber\\
\label{shalo}
\ea 
Therefore, given $m$ and a specific form of $\la\sigma|\hbox{v}|\ra$ 
associated with $a$ and $b$, equation (\ref{shalo}) provides a theoretical 
estimate of the entropy per particle of the neutralino halo gas that depends 
on the initial state given by $\xf$ in (\ref{eqxf}) and (\ref{sf}), on 
observable cosmological parameters $\Omega_0,\,h$ and on generic state 
variables associated to the halo structure.

\section{Theoretical vs. Empirical Entropies} 

If the neutralino gas in present halo structures would strictly satisfy MB 
statistics, the entropy per particle, $\shac$, in terms of $\rhohac=m\,\nhac$ 
and $\xhac=m\,c^2/(\kB\,\Tha_c)$, would follow from  the well known 
Sackur--Tetrode entropy formula \cite{Pathria}
\ba &&\shacmb = 
\frac{5}{2}
+\ln\left[\frac{m^4\,c^3}{\hbar^3\,(2\pi\,\xhac)^{3/2}\,\rhohac}\right]
\nonumber\\ &&= 94.42 +
\ln\left[\left(\frac{m}{\hbox{GeV}}\right)^4\,
\left(\frac{1}{\xha_c}\right)^{3/2}
\,\frac{\hbox{GeV/cm}^3}{\rhohac}\right].\nonumber\\
\label{s_halo}
\ea
Such a MB gas in equilibrium is equivalent to an isothermal halo if we 
identify \cite{BT} 
\ba 
\frac{c^2}{\xha} \ = \ \frac{\kB\,\Tha}{m} \ = \ \sigha^2,
\label{isot_MBa}
\ea
where $\sigha^2$ is the velocity dispersion (a constant for isothermal halos). 
However, as mentioned before, the assumption of MB statistics, in which 
entropy and energy are extensive quantities, does not apply for 
self--gravitating colissionless gases. Hence, an exactly isothermal halo is 
not a realistic model, not only because of these theoretical arguments, but 
also because its total mass diverges and its distribution function allows 
for infinite particle velocities (theoretically accessible in the velocity 
range of the MB distribution). Therefore, this case will not be considered 
any further. 

More realistic halo models follow from ``energy truncated'' (ET) distribution 
functions \cite{Padma3,BT,Katz1,Katz2,MPV} that assume a maximal ``cut off'' 
velocity (an escape velocity). Therefore, we can provide a convenient 
empirical estimate of the halo entropy, $\shac$, from the microcanonical 
entropy definition (\ref{mcsdef}) in terms of phase space volume, but 
restricting this volume to the actual range of velocities (\hbox{i.e.} 
momenta) accessible to the central particles, that is up to a maximal escape
velocity $v_e(0)$. It is reasonable to assume a relation of the form
\ba 
v_e^2(0) \ = \ 2\,|\Phi(0)| \ \simeq \ \alpha \, \sigha^2(0),
\label{alphas}
\ea
where $\Phi(r)$ is the newtonian gravitational potential. With the help of 
(\ref{alphas}) we have then
\ba &&\shacem \ \simeq \
\ln\left[\frac{m^4\,v_{e}^3}{(2\pi\hbar)^3\,\rhohac}\right]\nonumber\\ 
=&& 89.17
+\ln\left[\left(\frac{m}{\hbox{GeV}}\right)^4\,
\left(\frac{\alpha}{\xha_c}\right)^{3/2}
\,\frac{\hbox{GeV/cm}^3} {\rhohac}
\right],\nonumber\\
\label{SHALO}
\ea    
where we used $\xhac=c^2/\sigha^2(0)$ as in (\ref{isot_MBa}). As expected, 
the scalings of (\ref{SHALO}) are identical to those of (\ref{s_halo}).

In order to provide a numerical estimate for $\alpha$, we remark that by a 
suitable choice of a core radius and a total radius, $\RC$ and $\RT$, any 
realistic halo model can be accommodated to a generic galactic halo density
profile consisting of a region of constant density $0<r<\RC$, joined
continuously to an outer region $\RC<r<\RT$ in which $\rho$ decays as 
$r^{-2}$ and goes to zero for $r>\RT$.  If we further assume a velocity 
dispersion $\sigha^2(0)\simeq 2\,\vrot^2$, where $\vrot$ is the maximal 
observed rotation velocity, the parameter $\alpha$ defined in (\ref{alphas}) 
will be given by:
\begin{equation}
\alpha \ \simeq \ 2+ 4\,\ln \,\left(\frac{\RT}{\RC}\right)
\end{equation}
From this estimation, the minimum value of $\alpha$ occurs for the combination
of large $\RC$ and small $\RT$, with extreme values given (for a large spiral 
galaxy) by $10$ kpc and $100$ kpc, respectively. Conversely, the maximum value
of $\alpha$ corresponds to small $\RC$ and large $\RT$, with extreme values 
safely estimated as $1$ kpc and $300$ kpc, respectively. Considering these 
extreme values, we obtain the uncertainty range
\begin{equation} 
11.2 \ \alt \ \alpha \ \alt \ 24.8.
\label{alpha2}
\end{equation} 
Values of $\alpha$ outside these bounds are 
unrealistic~\cite{alpha1,alpha2,alpha3}, an assertion that can be also based 
on theoretical studies of dynamical and thermodynamical stability associated
with ET distribution functions \cite{Katz1,Katz2,cohn,MPV,HM,GZ,RST} and on 
observational data for normal spiral and LSB galaxies as well as galaxy 
clusters \cite{young,DBM,HG,FDCHA1,FDCHA2}. In any case, it must be noted 
that $\alpha$ is an integral feature of the halo, highly insensitive to the 
differential details of the density profile.

\section{Testing the Entropy Consistent Criterion}

The expressions for $\shac$ obtained from (\ref{shalo}) and (\ref{SHALO}) 
are not very useful, since they depend on dynamic halo variables, such as 
$\rhohac$ and $\xhac$, whose values depend on specific galactic systems and 
are subject to possibly large uncertainties. However, since both entropy 
estimates (\ref{shalo}) and (\ref{SHALO}) must correspond to the same 
quantity, $\shac$, the comparison of these independent estimates leads to 
the following constraint 
\ba 
\shacth &&= \ \shacem \quad \Rightarrow\nonumber\\ \xf && = \ 7.85 \ + \
\ln\left[\frac{(\alpha\,\xf)^{3/2}}{h^2\,\Omega_0}\,
\frac{m}{\hbox{GeV}}\right].
\label{constr}
\ea
which does not depend on halo variables other than $\alpha$, hence eliminating
the effects of potentially large uncertainties in the estimation of these 
quantities. In fact, the constraint (\ref{constr}) can be interpreted as the 
constraint on $\sff=5/2+\xf$ that follows from the condition 
$\shacth = \ \shacem$. Since we can use (\ref{eqOmega0}) and (\ref{eqYinf}) 
to eliminate $h^2\,\Omega_0$, the constraint (\ref{constr}) becomes a 
relation involving only $\xf,\,m,\,a,\,b,\,\alpha$. This constraint is 
independent of  (\ref{eqxf}), which is another (independent) expression for
$\sff=5/2+\xf$, but an expression that follows {\it{only}} from the 
neutralino annihilation processes. Therefore, the comparison between 
$\shacth$ and $\shacem$, leading to a comparison of two independent 
expressions for $\sff$, is not trivial but leads to  an 
``entropy consistency'' criterion that can be tested on suitable desired 
values of $m,\,a,\,b,\,\alpha$. This implies that a given dark matter 
particle candidate, characterized by $m$ and by specific annihilation 
channels given by $\xf$ through (\ref{eqxf}), will pass or fail to pass this 
consistency test independently of the details one assumes regarding the present
day dark halo structure. This is so, whether we conduct the consistency test by
comparing (\ref{eqxf}) and (\ref{constr}) or (\ref{shalo}) and (\ref{SHALO}). 
Since the matching of either (\ref{eqxf}) and (\ref{constr}) or (\ref{shalo}) 
and (\ref{SHALO}) shows a weak logarithmic dependence on $m$, the fulfillment 
of the  ``entropy consistency'' criterion identifies a specific mass range for 
each dark matter particle. This allows us to  discriminate, in favor or 
against, suggested dark matter particle candidates and/or annihilation 
channels by verifying if the standard abundance criterion (\ref{eqOmega0}) is
simultaneously satisfied for this range of masses. 

Since we can write (\ref{constr}) as:
\ba
\ln(h^{2} \Omega_0) \ = \ 7.85 - \xf 
+ \ln\left[(\alpha \,\xf)^{3/2} \, m \right],
\label{Result}
\ea
this constraint becomes a new estimate of the cosmological parameters 
$h^{2} \Omega_0$, given  in terms of a structural parameter of galactic dark 
matter halos, $\alpha$, the mass of the neutralino, $m$, and the temperature
of the neutralino gas at freeze out, $\xf$. This last quantity depends
explicitly not only on $m$, but also on its interaction cross section, and 
hence on the details of its phenomenological physics through (\ref{eqxf}). 
Notice that in reaching (\ref{Result}) we have used (\ref{mcsdef}) twice,  
both in estimating the change in entropy in going from the freeze out 
condition to the  present virialized halo, and in obtaining the actual 
empirical entropy in present day halos. However, in these two instances, 
(\ref{mcsdef}) was expressed in terms of two different sets of variables, 
involving $\Omega_0$ and $h$ in the former case (through the calculation of 
(\ref{eqnf}) and  (\ref{shalo})), and containing only halo parameters in the 
latter. In this way, equating the two expressions for the same quantity  
yields a constraint between the two sets of variables: equation (\ref{Result}).

At this point we consider values for the constants $a$ and $b$ that define the
interaction cross section of the neutralino, and use (\ref{Result}) to plot 
$\Omega_0$ as a function of $m$ in GeV's. Using $h=0.65$ and given the
uncertainty range of $\alpha$ in (\ref{alpha2}), we will obtain not a curve, 
but a region in the $\Omega_0-m$ plane. Considering first condition 
(\ref{Wboson}), corresponding to Higgsino--like neutralinos, leads to the 
shaded region in figure 1a. On this figure we have also plotted the relation 
which the abundance criterion (\ref{eqOmega0}) yields on this same plane. It 
is evident that within the observationally determined range of $\Omega_0$ 
(the horizontal dashed lines 0.2-0.4), there is no intersection between the 
shaded region and the abundance criterion curve. This implies that both 
criteria are mutually inconsistent, thus the possibility that Higgsino-like 
neutralinos make up both the cosmological dark matter and galactic dark 
matter can be ruled out.

Repeating the same procedure for mainly B--ino neutralinos, (\ref{sleptons}) 
yields figure 1b. In this case, we can see that the abundance criterion curve 
falls well within the shaded region defined by the entropy criterion. 
Although we can not improve on the mass estimate provided by the abundance 
criterion alone, the consistency of both criteria reveals the B-ino neutralino
as a viable option for both the cosmological and the galactic dark matter.

\begin{figure}
\centering
\includegraphics[height=15cm]{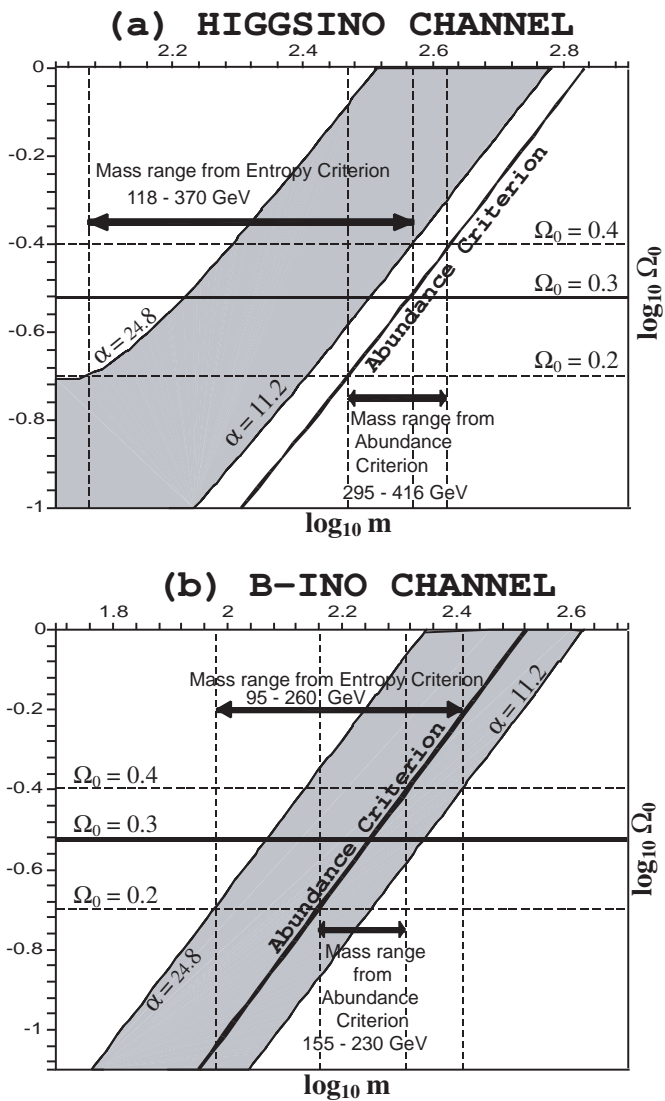}
%
%
\caption{Figures (a) and (b) respectively correspond to the Higgsino and B-ino 
channels. The shaded regions display $\Omega_0$ vs m from our entropy 
criterion (\ref{Result}) with the solid curve giving $\Omega_0$ from the 
cosmological abundance criterion (\ref{eqOmega0}), in all cases for $h=0.65$.
The horizontal dashed lines give current estimates of $\Omega_0=0.3\pm0.1$. 
It is evident that only the B-ino channels allow for a simultaneous fitting of 
both the abundance and the entropy criteria.}
\label{fig:2}       
\end{figure}

As noted above, the results of figures 1a and 1b are totally insensitive to 
the values of halo variables, $\xhac$ and $\rhohac$, used in evaluating 
(\ref{SHALO}) and (\ref{shalo}). Different values of these variables 
(say, for different halo structures) would only result in larger or smaller 
actual values of $\shac$ for each halo structure (subject to potentially large
uncertainties). Therefore, our results are not only insensitive to these 
uncertainties, but are also scale--invariant.  

\section{Conclusions} 

We have presented a robust consistency criterion that can be verified for any
annihilation channel of a given dark matter candidate proposed as the
constituent particle of the present galactic dark matter halos. Since we 
require that the empirical estimate $\shacem$ of present dark matter haloes 
must match the theoretical value $\shacth$, derived from the microcanonical 
definition and from freeze out conditions for  the candidate particle, the 
criterion is of a very general applicability, as it is largely insensitive 
to the details of the extremely complicated evolution of the neutralino gas 
from its freeze out era (hence, it is also insensitive to the structure 
formation scenario that might be assumed). Further, the details of the 
present day halo structure enter only through an integral feature of the 
dark halos, the central escape velocity, thus our results are also 
insensitive to the fine details concerning the central density and the 
various models describing the structure of dark matter halos. A crucial 
feature of this criterion is its direct dependence on the physical details 
({\it{i.e.}} annihilation channels and mass) of any particle candidate.  

Recent theoretical work by  E. A. Baltz {\it{et al.}} \cite{HEAT-TH} 
confirmed that neutralino annihilation in the galactic halo can produce 
enough positrons to make up for the excess of cosmic ray positrons 
experimentally detected by the HEAT collaboration \cite{HEAT1,HEAT2}. Baltz 
{\it{et al.}} concluded that for a boost factor $B_s \sim 30$ the 
neutralinos must be primarily B-inos with mass around 160 GeV. For a boost 
factor $30 < B_s < 100$, the gaugino--dominated SUSY models complying with 
all constraints yield neutralino masses in the range of 
$150\,\hbox{GeV} <  m_{{\tilde\chi^1_0}} < 400 \,\hbox{GeV}$. On the other 
hand, Higgsino dominated neutralinos are possible but only for 
$B_s \sim 1000$ with masses larger than 2 TeV. However, our results show 
this second option to be unlikely and are in agreement with the predictions 
that follow from \cite{HEAT-TH}, as we obtain roughly the same mass range 
for the B-ino dominated case (see figure 1b) and the Higgsino channel is 
shown to be less favored in the mass range lower than TeV's.       

We have examined the specific case of the lightest neutralino for the mostly
B-ino and mostly Higgsino channels. The joint application of the 
``entropy consistency'' and the usual abundance criteria clearly shows that 
the B-ino channel is favored over the Higgsino. This result can be helpful 
in enhancing the study of the parameter space of annihilation channels of 
LSP's in  MSSM models, as the latter only use equations (\ref{eqxf}) and 
(\ref{eqOmega0})--(\ref{eqYinf}) in order to find out which parameters
yield relic gas abundances that are compatible with observational constraints
\cite{Ellis,Report,Torrente,Roszkowski,Fornengo,Ellis2}. However, equations 
(\ref{eqxf}) and (\ref{eqOmega0})--(\ref{eqYinf}) by themselves are 
insufficient to discriminate between annihilation channels. A more efficient 
study of the parameter space of MSSM can be achieved by the joint usage of 
the two criteria, for example, by considering more general cross section 
terms (see for example \cite{Report}) than the simplified approximated forms 
(\ref{sleptons}) and (\ref{Wboson}). This work is currently in progress.\\   
         
\acknowledgements 

This work has been supported by DGAPA--UNAM grant No. 109001 and
CONACYT grants 137307--E and 139181--E.

\end{document}